# Dynamic Response of a fast near infra-red Mueller matrix ellipsometer

Lars M. S. Aas, Pål G. Ellingsen, Morten Kildemo and Mikael Lindgren

*Department of Physics, Norwegian University of Science and Technology (NTNU)*

N-7491 Trondheim, Norway

The dynamic response of a near infrared Ferroelectric Liquid Crystal based Mueller matrix ellipsometer (NIR FLC-MME) is presented. A time dependent simulation model, using the measured time response of the individual FLCs, is used to describe the measured temporal response. Furthermore, the impulse response of the detector and the pre-amplifier is characterized and included in the simulation model. The measured time-dependent intensity response of the MME is reproduced in simulations, and it is concluded that the switching time of the FLCs is the limiting factor for the Mueller matrix measurement time of the FLC-based MME. Based on measurements and simulations our FLC based NIR-MME system is estimated to operate at the maximum speed of approximately 16 ms per Mueller matrix measurement. The FLC-MME may be operated several times faster, since the switching time of the crystals depends on the individual crystal being switched, and to what state it is switched. As a demonstration, the measured temporal response of the Mueller matrix and the retardance of a thick liquid crystal variable retarder upon changing state is demonstrated.

## 1. Introduction

Fast Mueller matrix ellipsometers (MME) may have a wide range of applications in biology, medicine, and various monitoring and control systems, supplying additional polarimetric information in imaging and spectroscopic applications. A variety of high performance MMEs have been reported [1-8], but usually they are of considerable complexity and too slow for being used with CCD or CMOS stripe-spectrographs and imaging sensors. We are also considering the limit of non-reversible processes not covered by e.g. the pump-probe techniques used for ultrafast birefringence or dichroism measurements. Some authors have reported on fast Mueller matrix measurements [1-2, 9-10]. A typical solution uses a division of amplitude polarimeter and 4 detectors to work as the polarimeter or Polarization State Analyser (PSA). The modulation may be performed with a fast electro-optic [1, 3] or photo-elastic modulator [2]. However, such systems require a considerable increased complexity in instrumentation and signal processing. A recent development is the Snapshot Mueller matrix polarimeter based on spectral polarization coding, which enables Mueller matrix measurements within a µs [9-10]. However, this system does not allow for spectroscopic measurements, as the technique requires small spectral variations in the Mueller matrix of the sample. A rotating Fresnel prism retarder with a retardation of 132º across a wide spectral range (including UV) is probably the most optimal conditioned system around [5, 11]. However, the latter system is inherently slow due to the mechanical rotation of the retarders, thus making it less suitable for fast MME purposes.

Liquid crystal (LC) spatial light modulators have been developed over the past decades for a variety of applications. The LCs are of course most widely used in display applications where the amplitude mode of operation is desired, but phase control can be used in for example beam-steering and shaping [12-13], adaptive optics [14], optical tweezers [15], to mention a few. The Ferroelectric LCs are fast and allow for a rapid determination of the Mueller matrix. Several FLC based MME designs have been proposed, although the first proposal of a polarimeter system similar to the one reported on here, appears to have been by Gandorfer et al. [16]. However, as it is shown here, the FLCs are still the limiting factor for the switching speeds, in contrary to the findings reported recently [17-18]. The main disadvantage of liquid crystals, and in particular the current FLCs, is the well known degradation of the crystals upon ultra-violet radiation [19]. Applications of this technology in MME are currently thus limited to the visible and the infra-red. In this paper we investigate the dynamic response of a novel NIR-MME operating from 700 to 1600 nm. In a follow up paper we report on the features and applications of a FLC based NIR-MME imaging system based around the same design.

A key parameter in designing a Mueller matrix ellipsometer is to optimize the modulation or analyzer matrix of a polarization state generator (PSG) or a PSA in terms of their matrix condition numbers [5, 20-21]. In this respect, the Variable Liquid Crystal

Retarder (VLCR) based Mueller matrix ellipsometer can be truly optimal, but only for a single wavelength at the time. Furthermore, the switching of a VLCR is slow compared to the FLC variant. In fact, the dynamic possibilities of the fast near infrared Ferroelectric Liquid Crystal based MME are here demonstrated by characterization of the dynamic Mueller matrix of a VLCR during the switching.

## 2. Experimental details

The NIR-MME reported on here can be operated in spectroscopic mode from 750 nm to 1800 nm, using a grating monochromator and a single detector. A tungsten-halogen lamp (100 W) with a stabilized power supply was used for incoherent illumination. The system is alternatively operated by using a stable Laser diode operating at 980 nm, with maximum output power of 300 mW (found particularly useful for high speed measurements with short integration time).

A purpose built (by Elektron Manufaktur Mahlsdorf) extended InGaAs detector with a built in pre-amplifier, having an overall well designed flat frequency response and cut-off frequency at 150 kHz, was used in the spectroscopic mode.

A multifunctional NI-DAQ card with a maximum sampling rate of 1.25Ms/s (mega samples per second), is both used to acquire measurements and to control the FLCs. The MME is operated with in-house made Labview-based software. The initial FLC based MME design was chosen based on the design by Gandorfer et al. [16], and similar designs have later been reported elsewhere [7, 18]. In particular, Gandorfer et al. proposed a sequence of components such that the PSG and the PSA are composed of a Polarizer - fixed waveplate (WP1)- Ferroelectric Liquid Crystal (FLC1) –WP2 - FLC2. The remaining problem is then to determine the retardances and orientations of the waveplates and FLCs in order to optimize the condition number over the design spectral region. The PSA is in this initial design simply chosen with identical components in reverse order. Figure 1 shows a schematic drawing of the design.

The optical components in our design consist of a high contrast dichroic NIR polarizer, followed by a fixed "true zero-order" quarterwave retarder at 465 nm, a halfwave FLC at 510 nm, and a fixed halfwave retarder at 1008 nm, and finally a halfwave FLC at 1020 nm. The true zero-order waveplates where manufactured in quartz, while the FLCs were custom made versions of commercially available crystals (Displaytech Inc). The FLCs at 510 nm were the only elements fixed in the optimization, and originally chosen simply from cost considerations, while the fitted thicknesses of the remaining waveplates are the results of an optimization with respect to the condition number across the design spectral range.

The calibration samples were chosen as two NIR polarizers (similar to above) with azimuth orientation 0 and 90º relative to the first polarizer, and a zero-order waveplate at 1310 nm with azimuth orientation of the fast axis at 60º [18].

All the components were characterized individually, including retardance and azimuth orientation. The measurements were performed with collimated light and normal incidence to the optical polarization components. Simulation of the optical response of the full FLC based Mueller matrix ellipsometer was compared to the measured response, using the characteristic properties of each characterized component.

Measurements were performed on air, and on a custom made version of a commercially available VLCR (Meadowlark Inc.). The VLCR was designed to be a half-wave retarder at 2200 nm.

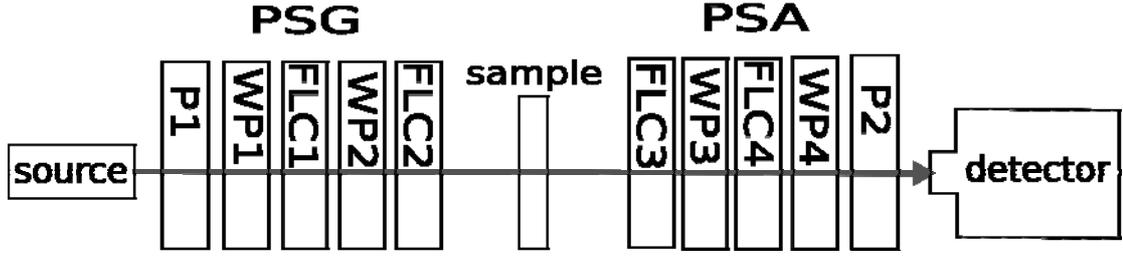

Figure 1. Layout of the FLC based NIR-MME, studied in this paper. The source is either a tungsten halogen white incoherent source or a 980 nm diode laser source. The FLC switching, timing and the signal acquisition are controlled by a computer. The optical components are ordered as a polarizer (P1), quartz waveplate (R1), FLC1 , waveplate (WP2), sample (e.g. air), FLC3, waveplate (WP3), FLC4, waveplate (WP4) and polarizer (P2). The detailed properties of the optical components are given in Table 1

### 3. Mueller matrix theory for FLC-Based MME

The MME is constructed by considering the Mueller matrix measurement theory proposed by Compain *et al.* [2, 20, 22-23]. In particular, the intensity matrix (**B**), which contains the 16 measured intensities, is given by the matrix product of the modulation matrix (**W**), the sample Mueller matrix (**M**) and the analyzer matrix (**A**)

$$\mathbf{B} = \mathbf{AMW} . \tag{1}$$

From the measured B matrix, and the known W and A matrices, the Mueller matrix can readily be calculated by matrix multiplication

$$\mathbf{M} = \mathbf{A}^{-1}\mathbf{B}\mathbf{W}^{-1} . \tag{2}$$

It is evident from basic matrix theory that **W** and **A**, for minimal error propagation, need to be as invertible as possible, i.e. as far from singular as possible. Specifically, it has been shown that the error in the intensity measurements, the calibration errors on **A** and **W**, are summed up into the errors in the resulting Mueller matrix as follows [5]

$$\frac{\|\Delta \mathbf{M}\|}{\|\mathbf{M}\|} \leq \kappa_\mathbf{A} \frac{\|\Delta \mathbf{A}\|}{\|\mathbf{A}\|} + \kappa_\mathbf{W} \frac{\|\Delta \mathbf{W}\|}{\|\mathbf{W}\|} + \kappa_\mathbf{A}\kappa_\mathbf{W} \frac{\|\Delta \mathbf{B}\|}{\|\mathbf{B}\|}, \tag{3}$$

where $\kappa_\mathbf{A}$ and $\kappa_\mathbf{W}$ are the condition numbers of **A** and **W**, defined in our work by the $L_2$ norm. This particular design method was originally proposed by Tyo [21], and later implemented and further developed by Compain *et al.* [1]. Compain *et al.* suggested an efficient and robust calibration technique (denoted the Eigenvalue Calibration Method) [20], which is also here implemented in order to determine the matrices **A** and **W**.

The modulation matrix **W** is constructed by defining each column vector as a unique Stokes vector generated by switching the two FLCs to one of their 4 different states. Let

$$\hat{\theta} = \{\theta_{1,FLC1}, \theta_{2,FLC1}, \theta_{1,FLC2}, \theta_{2,FLC2}, \theta_{1,FLC3}\theta_{2,FLC3}\theta_{1,FLC4}, \theta_{2,FLC4}\} \tag{4}$$

describe the set of azimuth angles of the FLCs. The two stable states $\{\theta_{1,j}, \theta_{2,j}\}$ of crystal $j$, are nominally separated by 45 degrees in this work. The modulation matrix is then given by

$$\mathbf{W} = [\vec{S}_1 \ \ \vec{S}_2 \ \ \vec{S}_3 \ \ \vec{S}_4], \tag{5}$$

where each stokes vector $\vec{S}_k$ is generated as:

$$\vec{S}_k = \vec{S}(\lambda, \hat{\theta}) = \mathbf{M}_{ret}\left(\theta_{i,FLC2}, \delta_{FLC2}(\lambda), T_{f,s}\right)\mathbf{M}_{ret}\left(\theta_{WP2}, \delta_{WP2}(\lambda), T_{f,s}\right) \times$$
$$\mathbf{M}_{ret}\left(\theta_{j,FLC1}, \delta_{FLC1}(\lambda), T_{f,s}\right)\mathbf{M}_{ret}\left(\theta_{WP1}, \delta_{WP1}(\lambda), T_{f,s}\right)\mathbf{M}_{pol}\begin{bmatrix}1 & 0 & 0 & 0\end{bmatrix}^t \tag{6}$$

where $\mathbf{M}_{ret}$ are the Mueller matrices of the retarders, oriented at a given angle $\theta$ with respect to the first polarizer (here used as a laboratory axis reference) with retardance $\delta$. $M_{pol}$ is the matrix of the polarizer horizontally aligned in our system. The index $i$ denotes the state of FLC2, and $j$ denotes the state of FLC1.

Similarly, the analyzer matrix is composed of the discrete analyzer states

$$\mathbf{A} = [\vec{A}_1 \ \ \vec{A}_2 \ \ \vec{A}_3 \ \ \vec{A}_4]^T. \tag{7}$$

Each analyzer state is then calculated as above, taking into account the reverse order of the optical components resulting in

$$\vec{A}_k = \vec{A}(\lambda, \hat{\theta}) = \begin{bmatrix}1 & 0 & 0 & 0\end{bmatrix}^t \mathbf{M}_{pol}\mathbf{M}_{ret}\left(\theta_{i,FLC4}, \delta_{FLC4}(\lambda)\right)$$
$$\times \mathbf{M}_{ret}\left(\theta_{FR4}, \delta_{FR4}(\lambda)\right)\mathbf{M}_{ret}\left(\theta_{j,FLC-3}, \delta_{FLC3}(\lambda)\right)\mathbf{M}_{ret}\left(\theta_{FR3}, \delta_{FR3}(\lambda)\right). \tag{8}$$

The index $i$ denotes the two stable states of FLC3, while $j$ denotes the two stable states of FLC4.

## 4. Results and Discussion

### *4.1 Static polarization properties of the optical components and the MME system*

The current version of our NIR-MME system has the characteristics summarized in Table 1, where all orientations of the polarizers, waveplates and the FLCs were individualy determined by separate Mueller matrix measurements in a commercial FLC-based visible MME instrument (Horiba Jobin Yvon). The analysis was performed by an implementation of the Lu-Chipman polar decomposition approach [24-26]. The dispersive retardances of each component in the NIR spectral region, was determined by a simple crossed polarizer set-up, as no commercial spectroscopic NIR-MME was available. Similarly, the transmittances of the fast and slow axes of the fixed wave-plates and the FLCs were determined by transmission measurements, by orienting the fast or the slow axis of these components with the transmission axis of the polarizers. The details of the measured dispersive retardances and transmission coefficients have been reported elsewhere [17]. For completeness some of the results are summarized here. The dispersive retardances where modelled by the following dispersion formula:

$$\Delta nd \approx \frac{A_{UV}\lambda}{\left(\lambda^2 - \lambda_{UV}^2\right)^{\frac{1}{2}}} - \frac{A_{IR}\lambda}{\left(\lambda_{IR}^2 - \lambda^2\right)^{\frac{1}{2}}} \tag{9}$$

where the retardances are given in nm at the wavelength $\lambda$. The constants $A_{UV}$, $A_{IR}$, $\lambda_{UV}$ and $\lambda_{IR}$ were determined by best fits to the data sets (Table 1).

The real static dispersive response of the MME may be quantified through the measurement of the condition number of the PSA and the PSG. The simulated response of the MME, using the values in Table 1, can be used to calculate the theoretical condition number. Figure 2 show both the measured and the simulated condition number of the PSG and the PSA. The system is in good correspondence with the simulations.

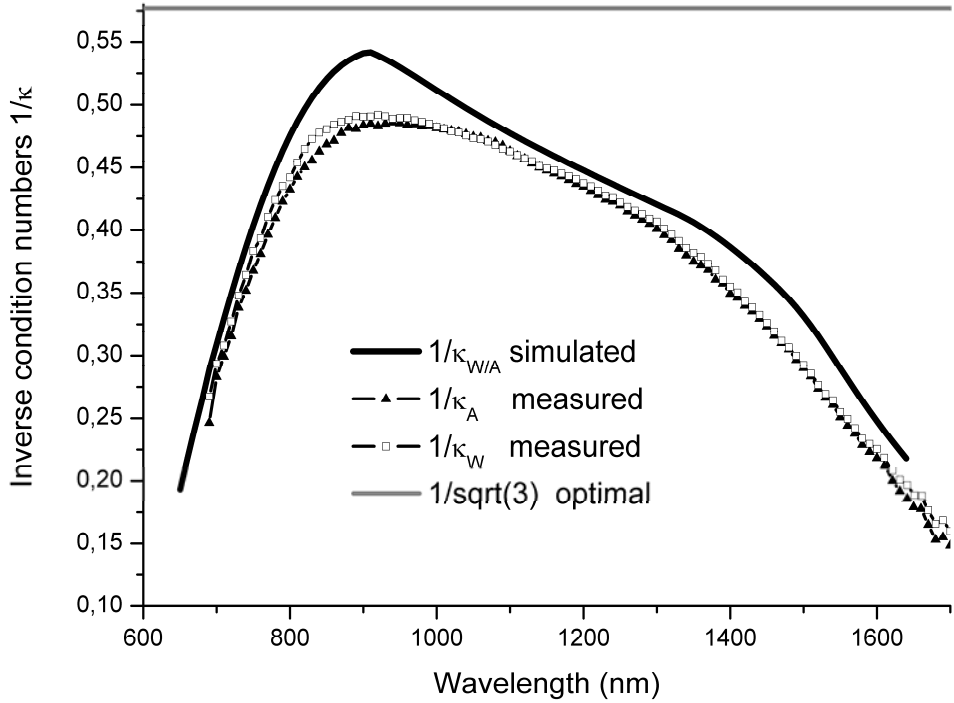

Figure 2. Simulated, measured and optimal inverse condition numbers for the Polarization State Generator and Analyzer, of the FLC-based Mueller matrix system, with the components tabulated in Table 1, and the setup in Figure 1.

Table 1. Static properties of polarization components in the order corresponding to Figure 1. The retardance is reported at a single wavelength, i.e. for WP1 then $\Delta nd = 116.25$ nm at $\lambda = 465$ nm. The wavelength dependent retardance is calculated from the fitted parameters $A_{IR}$, $A_{UV}$, $\lambda_{IR}$ and $\lambda_{UV}$. All azimuth orientations ($\theta$) of the fast axis of the stable state (denoted by "-"), are given with respect to the polarizers. Positive rotation of the azimuth axis is defined by anticlockwise rotation upon looking into the source [27].

|  | Retardance | $\theta$ (deg) | $\Delta\theta_{FLC}$ | $A_{UV}$ | $\lambda_{UV}$ [nm] | $A_{IR}$ | $\lambda_{UV}$ [μm] |
|---|---|---|---|---|---|---|---|
| *Polarizer* |  | 0 |  |  |  |  |  |
| *WP1* | λ/4@465nm | 129.2 |  | 110 | 134 | 50 | 11.16 |
| *FLC1* | λ/2@510 nm | 66.7(-) 110.7(+) | 44.0 | 202 | 280 | 0 |  |
| *WP2* | λ/2@1008 nm | 164.0 |  | 520 | 134 | 257 | 11.16 |
| *FLC2* | λ/2@1020 nm | 55.5(-) 100.9(+) | 45.4 | 505 | 283 | 0 |  |
| *FLC3* | λ/2@1020nm | 55.1 (-) 99.3 (+) | 44.2 | 505 | 283 | 0 |  |
| *WP3* | λ/2@1008 nm | 163.1 |  | 520 | 134 | 257 | 11.16 |
| *FLC4* | λ/2@510 nm | 68.5(-) 110.7(+) | 42.2 | 202 | 280 | 0 |  |
| *WP4* | λ/4@465nm | 129.4 |  | 110 | 134 | 50 | 11.16 |
| *Analyzer* |  | 0 |  |  |  |  |  |

## 4.2. Temporal optical response model for the fast FLC-based MME.

The deviation from a step-like temporal response of the recorded intensities in the measured **B** matrix arises as the system cannot switch immediately from one state to another, and that the photo-detector unit and its preamplifier might have a finite frequency response. Many future real-time applications of a MME could require the temporal characteristics of a certain polarization changing phenomenon. Moving objects, phase transitions, growth of a film or nanostructures, or variable orientation of molecules as in a liquid crystal are mentioned as a few examples.

Consider as an example a linear mechanical rotation of a waveplate (instead of using a FLC) by 45º, and the resulting intensity recorded by the detector as the waveplate is rotating. The switching of a FLC takes time, as the molecules in the LC need to be moved to their new orientations, which cause a nonlinear response of the recorded intensity as a function of the average angle of orientation. Furthermore, it is typical to drive such liquid crystals by an initial voltage spike, in order to enhance the switching frequency. As this involves a collective movement of the molecules making up the liquid crystals, the exact effect of the temporal response of the crystals in terms of its optical properties is uncertain.

To study the limitations in terms of the speed of a FLC based MME, the formalism in Section 3 (eqs. 3-7) was modified to include both the temporal characteristics of the rotating waveplates and the detector response. The results can readily be generalized to other discrete state MME designs, e.g. one based on electro-optic crystals. The measured temporal response can be compared to simulations, which can increase the understanding of the temporal characteristics of such systems, as will be exemplified below.

The generated Stokes vectors were modified in order to allow for the temporal characteristics of the time dependent angle of orientation between two stable states of the FLC, i.e. $\hat{\theta}(t)$, giving

$$\vec{S}_k = \vec{S}(\lambda, \hat{\theta}(t)). \tag{10}$$

By switching the system through the 16 states i.e., 4×4 different stable states of the FLCs, the recorded intensities are stored in the wavelength and time dependent, intensity matrix

$$\mathbf{B}(t,\lambda) = \mathbf{A}(t,\lambda)\mathbf{M}(\lambda)\mathbf{W}(t,\lambda).$$

If the temporal response of the sample Mueller matrix **M** is to be determined, it is evidently of utter importance that the time constants involved in the **A** and **W** matrices are much shorter than the ones involved in the sample Mueller matrix.

Figure 3a shows the temporal measured intensity response of the NIR-MME system. To keep the signal level high, the 980 nm diode laser source was used in these experiments. The sample was air, modelled as the identity matrix. It is a peculiar intensity response, with overshoots and undershoots before going to the stable states.

The FLCs are nominally reported to switch in 55 μs, and first it was investigated to what extent the overshoots could be due to the detector (and preamplifier) performance. This issue can easily be introduced into the simulations by introducing the impulse response of the preamplifier of the detector

$$\mathbf{B}(t,\lambda) = h(t) \otimes \mathbf{A}(t,\lambda)\mathbf{M}(\lambda)\mathbf{W}(t,\lambda), \tag{11}$$

were $h(t)$ is the detector impulse response, and $\otimes$ denotes the convolution. The detector pre-amplifier response was therefore characterized using the frequency modulation option of the diode laser and a lock-in amplifier. The frequency response analysis showed a flat response with a 80 dB/decade fall above 151 kHz. The results were fitted to the transfer function $H(s) = (1+sT)^{-4}$, where the time constant T = 1.05 μs/rad, giving the impulse response in the time domain of $h(t) = t^3 \left(6Te^{t/T}\right)^{-1}$. Although systematically included in the simulations below, the detector–preamplifier combination is many orders faster than the observed overshoots, and can be neglected in the analysis for Mueller matrix measurement frequencies

(i.e. 16 switches) below approximately 2.5 kHz. However, for even faster systems, the detector response will evidently affect the maximal Mueller matrix measurement frequency.

*4.2.1 Dynamic response of the FLCs.*

The FLCs can be regarded as waveplates with nominally two stable states, each separated by a rotation of the waveplate by 45°. In a more accurate model the switching between the stable states is obtained by rotating the molecular director axis of the molecules in the FLC by 180° in a cone around the normal axis of the molecular layer [19]. In order to understand the temporal limitations of the FLC-based MME it is most important to understand the collective dynamic response of the director axis. For simplicity, it is sufficient for our purposes to consider the approximation of crystal switching by assuming a simple rotation of a waveplate from the most stable state 1, to the second excited stable state 2. Let us denote "up -switching " when switching from the more stable to the excited state $(\theta_{1,j} \to \theta_{2,j})$, and "down-switching" when switching from the excited state to the more stable state $(\theta_{2,j} \to \theta_{1,j})$.

Since no fast Mueller matrix ellipsometer was available to measure the dynamic response of the switching of the crystals, a simple crossed polarizer setup was used. The FLCs were positioned with their fast axis at -20° with respect to the input polarizer. The measurements were performed at 980 nm, using the flat response InGaAs detector-preamplifier and the intensity was measured as a function of time. The estimated FLC orientation angle was solved by assuming the constant static retardance using Table 1, with the following equation:

$$I(t) = 1 - \cos^2 2\theta(t) - \cos\delta \sin^2 2\theta(t) \qquad (12)$$

Figure 4 shows the dynamic response of the crystals used in this study, using the latter method. The more correct model for the switching is obtained by precessing the tip of the molecular director in a cone around its axis of rotation. Then $\chi(t) = 0..\pi$ is related to the azimuth orientation angle by $\tan\theta(t) = \tan\varphi \cdot \cos\chi(t)$, with a fixed tilt angle of $\varphi(t) = \pi/8$. The latter precessing will also introduce a time-varying retardance. However, this model mainly modifies the centre of the switching curve, which is observed to be unaccurately determined in a crossed polarizer setup (see the discontinuous region around 22.5° in Figure 4). The details of the switching response obtained from Figure 4, is tabulated in Table 2, giving the transition times ($t_{up}$ and $t_{down}$) to the intensity first crosses the level of stable state, for the up and down switching, respectively. The transition times to steady state are given by $t_{ss-up}$ and $t_{ss-down}$, for the up and down switching respectively. A detailed inspection of the insets in Figure 4, shows that that the azimuth angle of the crystals appear to pass the stable state before settling, resulting in an overshoot, which is typically ranging from $\Delta\theta = 0.5 - 1°$, corresponding to a precession overshoot of $d\chi = 12 - 18°$. The switching times are found to vary from crystal to crystal, and the thicker crystals appear to have a 25 % longer switching time when going up and 35 % longer switching time when going down, compared with the thinner FLCs.

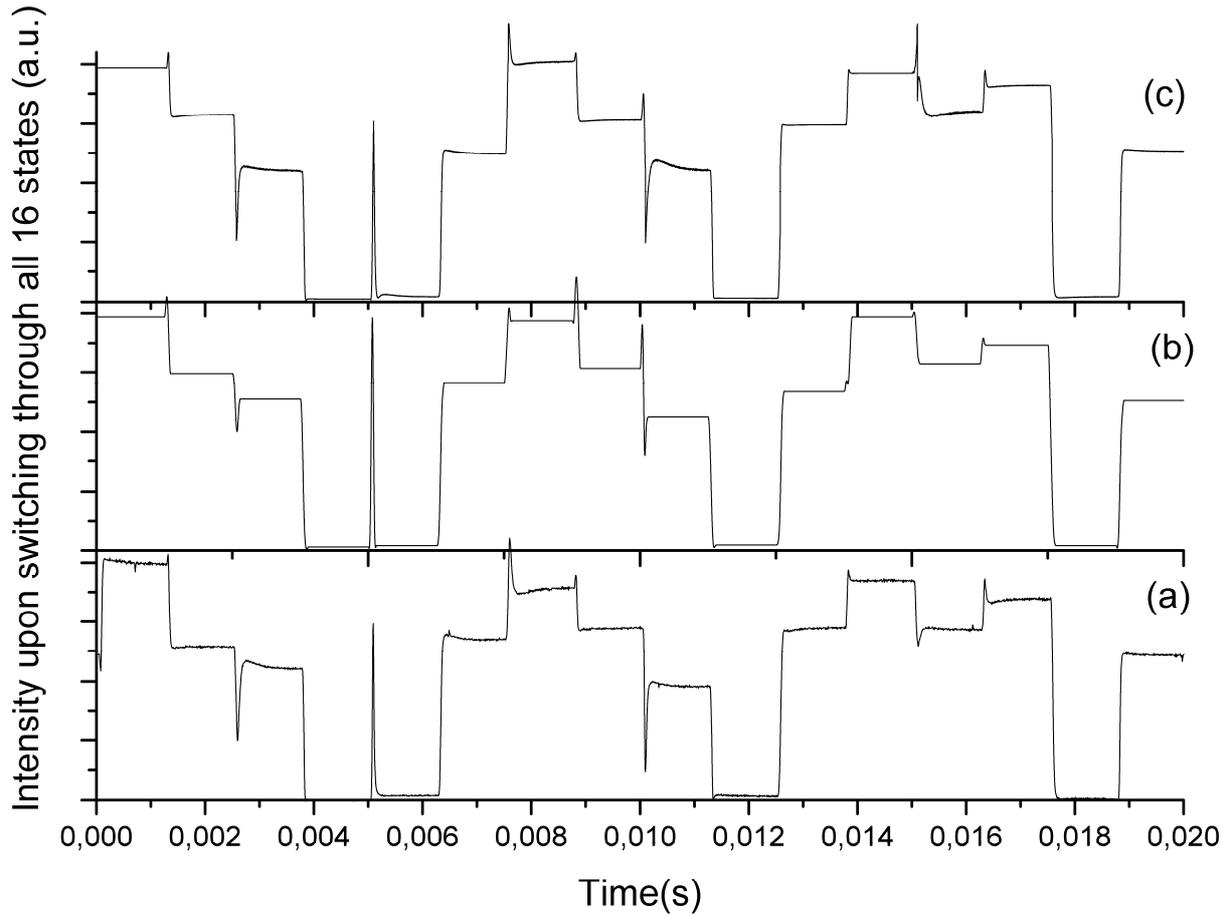

Figure 3. Measured and simulated temporal response of the recorded intensity upon switching through all 16 states of the FLCs. a) shows the measured intensity upon switching through all 16 states. b) shows the simulated intensity assuming a linear precession of the molecular director axis. c) shows the simulated intensity using the time dependent angular rotation from Figure 4. In both simulations the impulse response of the detector is included, and the switching times given by $t_{up}$ and $t_{down}$ in Table 2 is used.

*4.2.2 Simulation of the dynamic response of the FLC based MME*

The model for the optical intensities stored in **B**, was used to simulate a system that was, as far as possible, similar to the experimental set-up. In the first simulation, the crystals were imagined to switch semi-instantaneously between the states. It was observed that there were no overshoots, and the results were only slightly filtered by the response of the detector.

In a more realistic simulation, the crystals were linearly switched between the two stable states, using the measured up and down switching times tabulated in Table 2. The director axis was modeled to precess linearly in a cone towards the second state, i.e. by letting $\chi(t) = \pi t / t_{up,down}$. The retardance and the azimuth orientation were determined by using the standard Euler rotation of the dielectric tensor, which was assumed uniaxial with the extraordinary axis along the molecular director axis. Figure 3b shows the resulting simulated intensity response of the full MME upon switching through all the stable states. It is observed that the large overshoots and undershoots are well reproduced in the simulation, and we conclude that these are simply due to the time varying intensity illuminating the detector upon changing the Stokes vector as a function of time. A simple linear rotation of the azimuth angle $\theta(t)$ does not considerably modify the response. The simulated dynamic intensity response using the most advanced model for linear switching between the two states, Figure 3b, is not fully coinciding to the measured response shown in Figure 3a. The crystals have clearly a more complex response of the average azimuth angle, as was already demonstrated by the overshoots in the switching measurements of the FLCs in Figure 4. In the final simulation, the measured switching response of each crystal was therefore carefully introduced into the simulation in an identical manner to the system. The resulting simulated intensity obtained upon switching through the 16 states is shown in Figure 3c. It is observed that the simulation reproduces the measured response very well, and that the overall system response and its limitations may now be fully understood.

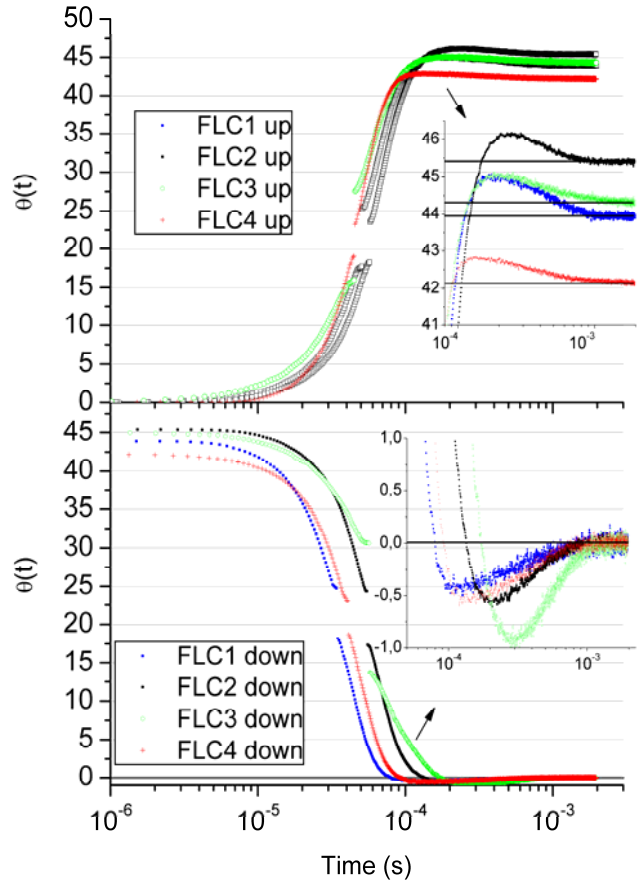

Figure 4. Measured temporal response of the azimuth angle $\theta(t)$ of the FLCs used in this study. The upper figure shows the up-switching of the crystals (more stable to less stable state), while the bottom figure shows the down-switching (less stable to more stable state). The temporal characteristics, such as $t_{up}$ (first crossover with steady state) and $t_{ss}$ (time to reach steady state) and the overshoot are tabulated in Table 2.

Table 2. Temporal switching characteristics of the FLC's used in the fast MME. The up switching and down switching times were used in the simulation in Figure 3.

|  | $t_{up}[\mu s]$ | $t_{ss\text{-}up}[\mu s]$ | overshoot[°] | $t_{down}[\mu s]$ | $t_{ss\text{-}down}[\mu s]$ | overshoot[°] |
|---|---|---|---|---|---|---|
| FLC1 (thin) | 113 | 800 | 1.00 | 78 | 850 | 0.44 |
| FLC2 (thick) | 149 | 950 | 0.72 | 133 | 1050 | 0.57 |
| FLC3 (thick) | 120 | 1000 | 0.54 | 173 | 1072 | 1.00 |
| FLC4 (thin) | 91 | 756 | 0.68 | 92 | 920 | 0.62 |

## 4.3 Application example: Dynamic characterization of a thick pneumatic Variable Liquid Crystal Retarder

As a demonstration of the operation of the fast FLC based NIR-MME, we have studied the temporal response of the Mueller matrix and the retardance of a thick Variable Liquid Crystal Retarder, as a function of the switching time, at 980 nm. The full Mueller matrix was measured for each time sample. Since the transition between states occur very quickly, the system was operated by sampling a Mueller matrix every 8 ms. For visualization purposes the measured Mueller matrix is approximated by

$$\mathbf{M}(t) = \begin{bmatrix} 1 & \vec{0}^T \\ \vec{0} & \hat{m}_R(t) \end{bmatrix},$$

where $\hat{m}_R(t)$ is the measured time dependent ($3\times 3$) retardance sub-matrix and $\vec{0} = [0,0,0]^T$ is the null vector. The resulting temporal retardance sub-matrix is shown in Figure 5. Figure 5 shows both $\hat{m}_R(t)$ observed by switching the VLCR by 277º (9 V to 2.5 V, $\Delta V$=6.5V) and by 90º (3 V to 2.5 V, $\Delta V$=0.5V). The azimuth orientation of the fast axis was -50º. The VLCR is observed to typically switch to its steady state in approximately 200 ms, independent of $\Delta V$. At maximum velocity, the VLCR switches with 8.6º per ms, the under-sampling will thus cause some errors in the measurements. The Lu-Chipman polar decomposition of the matrix gives the temporal evolution of the retardance, which is shown in Figure 6.

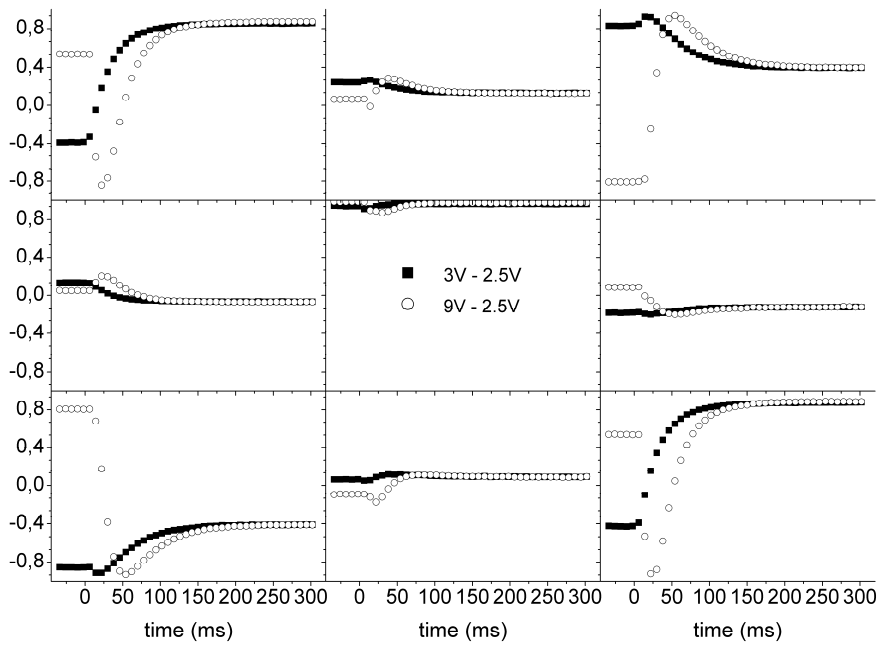

Figure 5. Temporal sub Mueller matrix ($\hat{m}_R(t)$) of a thick VLCR upon switching from 304º to 25º (circle) and 116º to 25º (square). The system was operated at 8 ms per Mueller matrix.

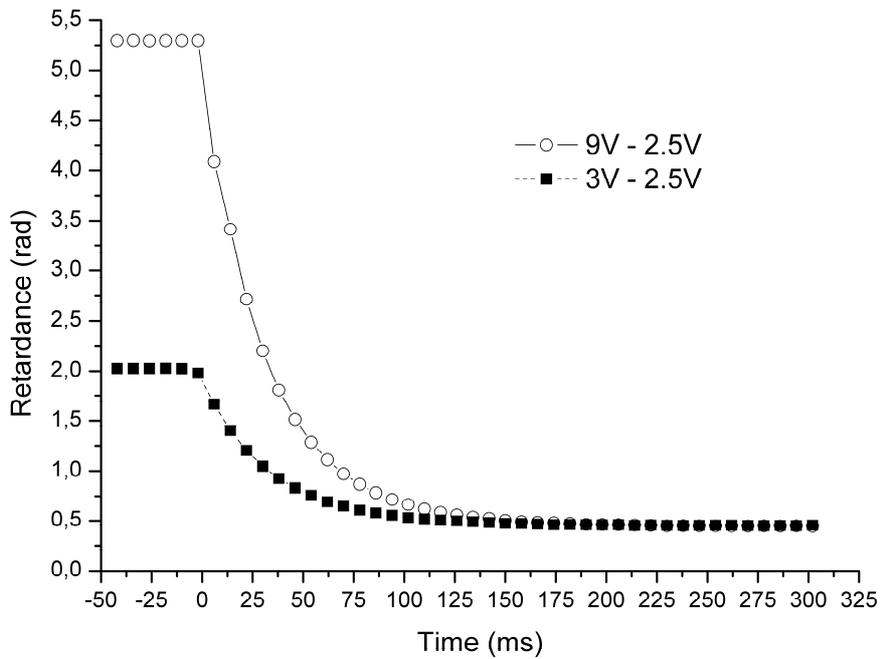

Figure 6. Dynamic response of the retardance (in radians) of a thick VLCR upon switching from 9 V to 2.5 V (circles), and 3V to 2.5 Volts (squares). The MME system was operated at 8 ms per Mueller matrix.

## 5. Summary and Conclusion

The dynamic response of a FLC based Mueller matrix Ellipsometer operating in the near infra-red has been measured and accurately modeled. The dynamic response of the intensity recorded on the detector is a function of the time dependent collective rotation of the molecular director axis and the impulse response of the detector. The simulations correspond well with the measurements, and it is concluded that using a flat and high frequency cut-off preamplifier and InGaAs detector does not limit the response, while the dynamic response of the LC molecules rotating towards steady state is the limiting factor for switching speed.

The switching of the crystals from one state to another result in an optical driving signal that may take various forms, and in the worst case has nearly the form of a delta-function. We have shown above, that this is a simple optical phenomenon that may be easily modeled by including a fast linear precession of the director axis of the molecules in the FLCs. Similar issues will be present in any fast operating MME. This response put certain requirements on the detector and preamplifier. However, the detector preamplifier designed for the current setup should be able to handle up to 2.5kHz repetition rates of Mueller matrix measurements. It is therefore the transient response of the switching of the FLCs that is the critical issue in the FLC-based MME. For accurate measurements, the steady state response, requires a minimum complete Mueller matrix measurement time of 15 ms. Further margin for data acquisition may also be required in order to obtain a good signal to noise ratio, resulting in 16 ms as a typical minimum requirement. The system may be operated faster, e.g. by only waiting for the $t_{up}$ and $t_{down}$, and allowing the additional measurement errors due to the overshoot. The system may then optimally be operated at a minimum measurement time of 2 ms in addition to acquisition times for each Mueller matrix.

As a demonstration the system has been successfully used to study the dynamic Mueller matrix during switching of a custom made thick variable liquid crystal, with a sampling of 8 ms per Mueller matrix.